# Scattered context-free linear orderings

Z. Ésik*

Department of Informatics
University of Szeged
Szeged, Hungary

December 22, 2017

## Abstract

We show that it is decidable in exponential time whether the lexicographic ordering of a context-free language is scattered, or a well-ordering.

## 1 Introduction

When the alphabet $A$ of a language $L \subseteq A^*$ is linearly ordered, $L$ may be equipped with the lexicographic order turning $L$ into a linearly ordered set. Every countable linear ordering may be represented as the lexicographic ordering of a language (over the two-letter alphabet). A (deterministic) context-free linear order is a linear ordering that can be represented as the lexicographic ordering of a (deterministic) context-free language. The study of context-free linear orderings has been initiated in [3]. In [4], it was shown that a well-ordering is deterministic context-free (or equivalently, definable by an algebraic recursion scheme) iff its order type is less than $\omega^{\omega^\omega}$. Then, in [5] it was shown that the Hausdorff rank of any deterministic context-free

*Partially supported by the project TÁMOP-4.2.1/B-09/1/KONV-2010-0005 "Creating the Center of Excellence at the University of Szeged", supported by the European Union and co-financed by the European Regional Fund, and the National Foundation of Hungary for Scientific Research, grant no. K 75249.

linear ordering is less than $\omega^\omega$. For an extension of these results to linear orderings definable by higher order recursion schemes we refer to [2].

Any monadic second-order definable property is decidable for deterministic context-free linear orders (given by $LR(1)$ grammars, say). This follows form a general decidability result for graphs in the pushdown hierarchy [6], more exactly from the "uniform version" of this result. In particular, it is decidable whether a deterministic context-free linear ordering is dense, or scattered, or a well-ordering. The results of [4, 5] implicitely give rise to practical algorithms for deterministic context-free languages. In contrast, as shown in [9], it is undecidable for a context-free linear ordering whether it is dense. The main results of this paper show that on the contrary, there is an exponential time algorithm to decide whether a context-free linear ordering is scattered, or a well-ordering. The fact that these properties are decidable for context-free linear orderings was first announced in [7].

## 2 Linear orderings

In this paper, by a linear ordering $L = (L, <)$ we shall mean a countable linear ordering. We will use standard terminology as in [11]. The isomorphism class of a linear ordering is its *order-type*.

A linear ordering $L$ is *dense* if it has at least two elements and for all $x, y \in L$, if $x < y$ then there is some $z$ with $x < z < y$. Up to isomorphism there are four (countable) dense linear orderings, the ordering of the rationals whose order-type is denoted $\eta$, possibly endowed with a least or greatest element, or both. A *scattered* linear order is a linear ordering that has no dense sub-order. A *well-ordering* is a linear ordering that has no sub-ordering isomorphic to the ordered set of the negative integers. Every well-ordering is scattered.

A linear ordering is *quasi-dense* if it is not scattered. It is well-known that any scattered sum or finite union of scattered linear orderings is scattered. Thus, if $I$ is a scattered linear ordering and for each $i \in I$, $L_i$ is a scattered linear ordering, then so is $\sum_{i \in I} L_i$. Moreover, if a linear ordering $L$ is the finite union of sub-orderings $L_i$, $i = 1, \ldots, n$, then $L$ is quasi-dense iff at least one the $L_i$ is quasi-dense.

Suppose that $A$ is an alphabet whose letters are ordered by $a_1 < \ldots < a_k$. Then we define the *strict order* $<_s$ on the set of words $A^*$ by $u <_s v$ iff $u = xa_iy$ and $v = xa_jy'$ for some $x, y, y' \in A^*$ and letters $a_i$ and $a_j$ with $a_i < a_j$. The *prefix order* is defined by $u <_p v$ iff $u$ is a proper prefix of $v$. The strict order and the prefix order are partial orders. The *lexicographic order* $<_\ell$ is the union of the two, so that $x <_\ell y$ iff $x <_s y$ or $x <_p y$. Clearly, $(A^*, <_\ell)$ is a linear ordering.

If $L \subseteq A^*$ then $(L, <_\ell)$ is a linear ordering, called the *lexicographic ordering of $L$*. We call $L$ dense, scattered or well-ordered if $(L, <_\ell)$ has the appropriate property. When $L$ is a (deterministic) context-free language, we call $(L, <_\ell)$, and sometimes any linear ordering isomorphic to $(L, <_\ell)$ a *(deterministic) context-free linear ordering*. Every (deterministic) context-free linear ordering is isomorphic to the lexicographic ordering of a (deterministic) context-free language over the alphabet $\{0, 1\}$, ordered by $0 < 1$. Indeed, when $L \subseteq A^*$ and $A$ has $k$ letters $a_1 < \ldots < a_k$, say, then we may encode each letter $a_i$ with a binary word $h(a_i)$ of length $\lceil \log k \rceil$ over $\{0, 1\}$ so that $h(a_i) <_\ell h(a_j)$ whenever $a_i < a_j$, then $(L, <_\ell)$ is isomorphic to $(h(L), <_\ell)$.

When $L \subseteq \{0, 1\}^*$ then $T(L)$ is the binary tree whose vertices are the words in the prefix closure of $L$. $T(L)$ is nonempty if $L$ is nonempty. A vertex $y$ is a descendant of vertex $x$ if $x$ is a prefix of $y$. The following fact is quite standard:

**Proposition 2.1** *Suppose that $L \subseteq \{0,1\}^*$ and consider the corresponding tree $T(L)$. Then $L$ is quasi-dense iff the full binary tree has an embedding in $T(L)$.*

*Proof.* Let $L_0$ be the regular prefix language $(00+11)^*01$ whose lexicographic ordering has order type $\eta$, and consider the tree $T(L_0)$. If the full binary tree embeds in $T(L)$, then so does $T(L_0)$. Consider an embedding of $T(L_0)$ in $T(L)$ which maps each vertex $x$ of $T(L_0)$ to a vertex $h(x)$ of $T(L)$. For each leaf $x$ of $T(L_0)$ select a leaf $v_x$ of $T(L)$ which is a descendant of $h(x)$. The words $v_x$ form a dense subset of $L$ with respect to the lexicographic order. (Note also that any two words $v_x$ are actually related by the strict order.)

For the reverse direction, suppose that $L$ is quasi-dense. Let us color a vertex $x$ of $T(L)$ blue if $x \in L$. Call a vertex $x$ of $T(L)$ *appropriate* if the blue vertices of the subtree $T_x$ rooted at $x$ form a quasi-dense linear ordering with respect to the lexicographic order. If $x$ is appropriate, then it has at least two proper descendants $y$ and $z$ which are appropriate vertices with $y <_s z$. Indeed, $x$ has a proper descendant $x'$ such that both the set of blue vertices $y'$ of $T_x$ with $y' <_s x'$ and the set of blue vertices $z'$ of $T_x$ with $x' <_\ell z'$ form quasi-dense linear orderings with respect to the lexicographic order. Suppose that $x' = xu$, where $u$ is a nonempty word. Then one of the vertices $xv0$ where $v1$ is a prefix of $u$ is appropriate, as is one of the vertices $xv1$ and $x'$, where $xv0$ is a prefix of $x$. Let $y$ and $z$ be these vertices.

Thus, starting from the root of $T(L)$, we can construct a set $V$ of appropriate vertices such that each $x \in V$ has two (proper) descendants $y$ and $z$ in $V$ with $y <_s z$. The vertices in $V$ determine an embedding of the full binary tree in $T(L)$. □

**Proposition 2.2** *Suppose that $L, L' \subseteq \{0,1\}^*$. If $(L, <_\ell)$ and $(L', <_\ell)$ are both scattered, then so is $(LL', <_\ell)$.*

*Proof.* We will prove that if $(LL', <_\ell)$ is quasi-dense, then one of $(L, <_\ell)$ and $(L', <_\ell)$ is quasi-dense. Assuming that $LL'$ is quasi-dense, $T(LL')$ has an embedded copy $T_0$ of the full binary tree. Let us color a vertex $u$ of $T(LL')$ blue if $uv \in L$ for some $v \in \{0,1\}^*$, i.e., when $u$ has a descendant in $L$. There are two cases to consider, either each subtree of $T_0$ contains a blue vertex, or there is a subtree of $T_0$ having no blue vertex.

*Case 1.* Suppose that each subtree of $T_0$ contains a blue vertex. Then each vertex of $T_0$ is colored blue, so that $L$ is quasi-dense.

*Case 2.* Suppose that $T_0$ contains a subtree having no blue vertex. Let $T_1$ denote such a subtree and let $u$ denote the root of $T_1$. Let $u_0, \ldots, u_k$ be all the (proper) prefixes of $u$ that are in $L$. Now let us color each vertex $x$ of $T_1$ with the set of all integers $i$, $0 \leq i \leq k$, such that $x$ has a descendant in $T(LL')$ which is a word in $u_i L'$. Then each vertex $x$ of $T_1$ is labeled by a nonempty subset of the set $\{0, \ldots, k\}$, and if $x'$ is a descendant of $x$ in $T_1$, then the label of $x'$ is included in the label of $x$. Let $H$ be a minimal set that appears as the label of a vertex $v$ of $T_1$. Then all descendants of $v$ in $T_1$ are labeled $H$. Thus, if $i \in H$, then the full binary tree embeds in $T(u_i L')$ and thus in $T(L')$, so that $L'$ is quasi-dense. □

## 3 Scattered context-free linear orderings

In this section, we assume that $G = (N, \{0,1\}, P, S)$ is a context-free grammar with nonterminal alphabet $N$, terminal alphabet $\{0,1\}$, rules $P$ and start symbol $S$ that contains no useless nonterminals or $\epsilon$-rules. Moreover, we assume that $G$ is left-recursion free and that $L(G)$ is not empty. These can be assumed for the results of the paper, since there is an easy polynomial time transformation of a context-free grammar to a grammar over the alphabet $\{0, 1\}$ that generates an isomorphic language (with respect to the lexicographic order) not containing $\epsilon$, and each grammar not generating the empty word can be transformed in polynomial time into an equivalent grammar that contains no useless nonterminals or $\epsilon$-rules or any left-recursive nonterminal. See [1, 8].

We let $X, Y, Z$ (sometimes decorated) denote nonterminals, $u, v, w, x, y$ terminal words in $\{0, 1\}^*$, and we let $p, q, r$ denote words in $(N \cup \{0, 1\})^*$. For every word $p$, we denote by $L(p)$ the set of all words $w \in \{0, 1\}^*$ with $p \Rightarrow^* w$. Thus, the language $L(G)$ generated by $G$ is $L(S)$. The length of $p$ is denoted $|p|$.

For nonterminals $X$ and $Y$ we define $Y \preceq X$ iff there exist $p, q$ with $X \Rightarrow^* pYq$, and we define $X \approx Y$ if both $X \preceq Y$ and $Y \preceq X$ hold. When $X \approx Y$, we say that $X$ and $Y$ belong to the same *strong component*. When $Y \preceq X$ but $X \not\approx Y$, we also write $Y \prec X$. The *height* of a nonterminal $X$ is the length $k$ of the longest sequence $Y_1 \prec \ldots \prec Y_k = X$. When $\mathcal{C}$ is a strong component and $X \in \mathcal{C}$ has height $k$, we also say that $\mathcal{C}$ has height $k$.

A *primitive word* is a nonempty word that is not a proper power. For elementary properties of primitive words we refer to [10].

**Theorem 3.1** *The following conditions are equivalent.*

1. $(L(G), <_\ell)$ *is a scattered linear ordering.*

2. *There exist no nonterminal $X$ and words $u, v \in \{0, 1\}^*$ such that neither $u$ is a prefix of $v$ nor $v$ is a prefix of $u$, moreover, $X \Rightarrow^* uXp$ and $X \Rightarrow^* vXq$ hold for some $p, q$.*

3. *For each recursive nonterminal $X$ there is a primitive word $u_0 = u_0^X$ such that whenever $X \Rightarrow^+ wXp$ then $w \in u_0^+$.*

4. *For each strong component $\mathcal{C}$ containing a recursive nonterminal there is a primitive word $u_0 = u_0^\mathcal{C}$, unique up to conjugacy, such that for all $X, Y \in \mathcal{C}$ there is a (necessarily unique) conjugate $v_0$ of $u_0$ and a proper prefix $v_1$ of $v_0$ such that if $X \Rightarrow^+ wYp$ for some $w \in \{0, 1\}^*$ and $p \in (N \cup \{0, 1\})^*$ then $w \in v_0^* v_1$.*

*Proof.* It is easy to prove that the first condition implies the second. Suppose that $L(G)$ is scattered and let $X \Rightarrow^* uXp$ and $X \Rightarrow^* vXq$. If neither $u$ is a prefix of $v$ nor $v$ is a prefix of $u$, then $u$ and $v$ are nonempty and comparable with respect to the strict order, say $u <_s v$. Suppose that $S \Rightarrow^* wXp$. The vertices $w(u+v)^*$ determine an embedding of the full binary tree in $T(L)$. Thus, by Proposition 2.1, $L$ is quasi-dense, a contradiction. Thus, either $u$ is a prefix of $v$ or vice versa.

Suppose now that the second condition holds. We prove that the third condition also holds. Let $X$ be a recursive nonterminal and suppose that $X \Rightarrow^+ uXp$. Then $u$ is nonempty (since $G$ is left recursion free) and thus has a primitive root $u_0$. We claim that whenever $X \Rightarrow^+ wXq$ then $w$ is a power of $u_0$. Indeed, if $X \Rightarrow^+ wXq$ then $w$ is also nonempty and thus there exist $m, n > 0$ with $|u^n| = |w^m|$. Since $X \Rightarrow^+ u^n X p^n$ and $X \Rightarrow^+ w^m X q^m$, it follows that $u^n = w^m$, so that $u_0$ is also the primitive root of $w$.

Next we prove that the third condition implies the fourth. So assume that the third condition holds. Note that if a strong component contains a recursive nonterminal, then all nonterminals in that strong component are recursive.

**Lemma 3.2** *Suppose that $X, Y$ are different recursive nonterminals that belong to the same strong component. Then $u_0^X$ and $u_0^Y$ are conjugate.*

*Proof.* Since $X, Y$ belong to the same strong component, there exist $x, y$ and $p, q$ with $X \Rightarrow^+ xYp$ and $Y \Rightarrow^+ yXq$. Thus, $X \Rightarrow^+ xyXqp$ and $Y \Rightarrow^+ yxYpq$. Thus, $xy$ is a power of $u_0^X$ and $yx$ is a power of $u_0^Y$. Since $xy$ and $yx$ are conjugate and $u_0^X$ and $u_0^Y$ are primitive, this is possible only if $u_0^X$ and $u_0^Y$ are conjugate. □

Using the lemma, we now complete the proof of the fact that the third condition implies the fourth.

Suppose that the strong component $\mathcal{C}$ contains a recursive nonterminal and $X_0 \in \mathcal{C}$. Let $u_0^\mathcal{C} = u_0^{X_0}$. For the sake of simplicity, below we will just write $u_0$ for this word. Let $X, Y \in \mathcal{C}$ with $X \Rightarrow^+ wYp$ and $Y \Rightarrow^* xXq$, where $w, x, p, q$ are appropriate words, so that $X \Rightarrow^+ wxXqp$. By Lemma 3.2 we have that $wx$ is a power of a primitive word $v_0$ which is a conjugate of $u_0$. It is clear that $v_0$ is unique. Also, $w = v_0^n v_1$ for some $n \geq 0$ and some proper prefix $v_1$ of $v_0$.

We still need to show that if $X \Rightarrow^+ w'Yp'$ for some $w'$ and $p'$, then $w'$ can be written as $v_0^m v_1$ for some $m$. But in this case $X \Rightarrow^+ w'xXqp'$ and $w'x$ is a power of $v_0$. Since the length of $w'$ is congruent to the length of $w$ modulo the length of $v_0$, it follows that $w' = v_0^m v_1$ for some $m \geq 0$. This ends the proof of the fact that the third condition implies the fourth.

Suppose finally that the fourth condition holds. Then clearly, the third condition also holds. We want to prove that $L(G)$ is scattered. To this end, we establish several preliminary facts.

**Definition 3.3** *Suppose that $X$ is a recursive nonterminal and let $u_0 = u_0^X$. For each $n \geq 0$ and prefix $ui$ of $u_0$, where $i = 0, 1$, let $L(X, n, ui)$ denote the set of all words of the form $u_0^n u\bar{i}w$ in $L(X)$, where $w \in \{0, 1\}^*$ and $\bar{i} = 1$ iff $i = 0$.*

Let $X$ be a recursive nonterminal. The following facts are clear. (Below we continue writing $u_0$ for $u_0^X$.)

**Proposition 3.4** *Each word in $L(X)$ is either in $L(X, n, ui)$ for some $n \geq 0$ and prefix $ui$ of $u_0$, or is a word of the form $u_0^n u$ where $n \geq 0$ and $u$ is a proper prefix of $u_0$.*

**Proposition 3.5** *For each $n \geq 0$ and prefix $ui$ of $u_0$ there is only a finite number of left derivations*

$$X \Rightarrow_\ell^* wYp \Rightarrow_\ell u_0^n u\bar{i}q \tag{1}$$

*such that $u_0^n u\bar{i}$ is not a prefix of $w$.*

Let us denote by $F(X, n, ui)$ the finite set of all words $q$ that occur in derivations (1).

**Proposition 3.6** *If $Y$ is a nonterminal that occurs in a word $q \in F(X, n, ui)$ for some $n \geq 0$ and prefix $ui$ of $u_0$, then $Y \prec X$.*

*Proof.* Suppose that (1) is a left derivation and $Y$ occurs in $q$, so that $q = q_1 Y q_2$ for some $q_1, q_2$. If $X \approx Y$ then there exist some $r_1, r_2$ with $Y \Rightarrow^* r_1 X r_2$. Thus, $q = q_1 Y q_2 \Rightarrow^* q_1 r_1 X r_2 q_2$. Let $v$ denote a terminal word with $q_1 r_1 \Rightarrow^* v$. Then we have

$$X \Rightarrow^* u_0^n u\bar{i}q \Rightarrow^* u_0^n u\bar{i}vXr_2q_2.$$

Since $u_0^n u\bar{i}v$ is not a power of $u_0$, this contradicts the third condition. □

We now complete the proof of Theorem 3.1.

Let $X$ be a nonterminal. We prove the following fact: If $(L(Y), <_\ell)$ is scattered for all nonterminals $Y$ whose height is less than the height of $X$, then $(L(X), <_\ell)$ is scattered.

If $X$ is not a recursive nonterminal, then the height of each nonterminal appearing on the right side of a rule $X \to p$ is less than the height of $X$. Thus $L(X)$ is the finite union of all languages $L(p)$ where $X \to p$ is in $P$. By the induction hypothesis and Proposition 2.2, each linear ordering $(L(p), <_\ell)$ is scattered. Since any finite union of scattered linear orderings is scattered, $(L(X), <_\ell)$ is also scattered.

Suppose now that $X$ is recursive. Then by Proposition 3.4,

$$L(X) = L_0 \cup \bigcup_{n \geq 0,\ ui} L(X, n, ui)$$

where $ui$ ranges over the prefixes of $u_0 = u_0^X$ and each word of $L_0$ is of the form $u_0^n v$ for some $n \geq 0$ and some proper prefix $v$ of $u_0$. It is clear that $L_0$ is scattered (in fact, either a finite linear ordering or an $\omega$-chain). Thus, it suffices to show that

$$(\bigcup_{n \geq 0,\ ui} L(X, n, ui), <_\ell)$$

is scattered. But this linear ordering is isomorphic to the ordered sum

$$\sum_{n \geq 0,\ u1} L(X, n, u1) + \sum_{n \leq 0,\ u0} L(X, -n, u0)$$

where in the first term $u1$ is a prefix of $u_0$ and in the second term $u0$ is a prefix of $u_0$. Since a scattered sum of scattered linear orderings is scattered, it remains to show that each $L(X, n, ui)$ is scattered. But by Proposition 3.5, for each $n$ and $ui$, $L(X, n, ui)$ is a finite union of languages of the form $u_0^n u\bar{i}L(q)$ where $q$ contains only nonterminals of height strictly less than the height of $X$. Thus, by the induction hypothesis and Proposition 2.2, each such language is scattered. Since any finite union of scattered linear orderings is scattered, it follows that $L(X, n, ui)$ is scattered. This ends the proof of the fact that the fourth condition implies the first. The proof of Theorem 3.1 is complete. □

**Theorem 3.7** *$L(G)$ is well-ordered iff $L(G)$ is scattered and there is no recursive nonterminal $X$ such that $L(X)$ contains a word $w$ such that $u_0^n <_s w$ for some $n$, where $u_0 = u_0^X$.*

*Proof.* Note that the extra condition is equivalent to that for all recursive nonterminals $X$ and for any prefix $u0$ of $u_0 = u_0^X$, the language $L(X, n, u0)$ is empty. Now by repeating the last part of the proof of Theorem 3.1, it follows that under this condition, if $L(G)$ is scattered, then $L(X)$ is well-ordered for all $X$. One uses the well-known fact that if a linear order is a finite union of well-orderings, then it is also a well-ordering, and that a well-ordered sum of well-orderings is well-ordered.

On the other hand, if the extra condition is not satisfied for the recursive nonterminal $X$, then $L(X)$ is not well-ordered. For suppose that $L(X, n, u0)$ contains the word $u_0^n u1x$. We know that there is some $m \geq 1$ and some $w$ with $X \Rightarrow^+ u_0^m X w$. Thus, the words $u_0^{km} u_0^n u1xv^{km}$ for $k = 0, 1, \ldots$ form a strictly decreasing sequence in $L(X)$. We conclude by noting that if $L(X)$ is not well-ordered for some $X$, then $L(G)$ is not well-ordered either, since $G$ contains no useless nonterminals. □

At this point, we are already able to show that it is decidable whether $L(G)$ is scattered, or well-ordered.

**Corollary 3.8** *There exists an algorithm to decide whether $L(G)$ is scattered.*

*Proof.* As before, we may assume that $G = (N, \{0, 1\}, P, S)$ contains no useless nonterminals or $\epsilon$-rules. Moreover, we may assume that $G$ is left-recursion free and that $L(G)$ is not empty. By Theorem 3.1 we know that $L(G)$ is scattered iff for each recursive nonterminal $X$ there is a primitive word $u_0$ such that whenever $X \Rightarrow^+ wXp$ then $w \in u_0^+$. We are going to test this condition. Given a recursive nonterminal $X$ in the strong component $\mathcal{C}$, we find a word $u$ such that $X \Rightarrow^+ uXp$ for some $p$. Clearly, $u \neq \epsilon$. Let $u_0$ denote the primitive root of $u$. Then consider the following grammar $G_X$. The nonterminals are the nonterminals of $G$ together with the nonterminals $\overline{Y}$, where $Y \in \mathcal{C}$. The rules are those of $G$ together with the rules

$$\overline{Y} \to p\overline{Z}$$

such that $Y, Z \in \mathcal{C}$ and there is some $q$ with $Y \to pZq \in P$. There is one more rule, $\overline{X} \to \epsilon$. Let $\overline{X}$ be the start symbol. Then $L(G_X) \subseteq u_0^*$ iff for all $w$ such that $X \Rightarrow^+ wXp$ for some $p$ in $G$, it holds that $w \in u_0^+$. Now $L(G_X) \subseteq u_0^*$ iff the intersection of $L(G_X)$ with the complement of $u_0^*$ is empty, which is decidable. □

**Corollary 3.9** *There exists an algorithm to decide whether $L(G)$ is well-ordered.*

*Proof.* The extra condition introduced in Theorem 3.7 can be effectively tested, since it says that for each recursive nonterminal $X$, the intersection of $L(X)$ with the regular language of all words of the form $u_0^n u1x$, where $n \geq 0$ and $u0$ is a prefix of $u_0$, is empty. □

## 4 Decidability in exponential time

In this section, we give somewhat more efficient algorithms. First we need some preparation.

Suppose that $u_0 \in \{0,1\}^*$ is a fixed primitive word, and consider the set $\mathcal{S}$ of all pairs $(x_1, x_2)$, where $x_1$ is a proper suffix of $u_0$ and $x_2$ is a proper prefix of $u_0$. In particular, $(\epsilon, \epsilon) \in \mathcal{S}$. With each $(x_1, x_2) \in \mathcal{S}$ we associate the language $L(x_1, x_2) = x_1 u_0^* x_2$, if $|x_1 x_2| < |u|$, and $L(x_1, x_2) = x_1 u_0^* x_2 + z$ where $z$ is the suffix of $x_1 x_2$ obtained by removing its prefix of length $|u_0|$, if $|x_1 x_2| \geq |u_0|$. (Note that the prefix of length $|u_0|$ of $x_1 x_2$ is a primitive word which is a conjugate of $u_0$.) We call a word $w$ *legitimate* if it belongs to $L(x_1, x_2)$ for some $(x_1, x_2) \in \mathcal{S}$. Clearly, a word is legitimate iff it is a subword of some power of $u_0$ iff it is in $v_0^* z$ for some conjugate $v_0$ of $u_0$ and some necessarily unique proper prefix $z$ of $v_0$. Moreover, for each $(x_1, x_2) \in \mathcal{S}$ there is a unique conjugate $v_0$ of $u_0$ and a unique proper prefix $z$ of $v_0$ with $L(x_1, x_2) = v_0^* z$. It follows from this fact that any two languages $L(x_1, x_2)$ and $L(y_1, y_2)$ for $(x_1, x_2) \neq (y_1, y_2)$ in $\mathcal{S}$ are either disjoint or have a single common element which is a proper subword of $u_0$. In particular, for any legitimate word $u$ with $|u| \geq |u_0|$ there is a unique $(x_1, x_2) \in \mathcal{S}$ with $u \in L(x_1, x_2)$.

It is also clear that any subword of a legitimate word is legitimate, and if $u \in L(x_1, x_2)$, say, and $v$ is obtained from $u$ by removing a subword of length $|u_0|$, then $v$ is legitimate with $v \in L(x_1, x_2)$. Also, if $v$ is obtained by duplicating a subword of $u$ of length $|u_0|$ then $v$ is legitimate with $v \in$ $L(x_1, x_2)$. Moreover, when $u, v \in L(x_1, x_2)$, then $|u|$ is congruent to $|v|$ modulo $|u_0|$.

Let $(x_1, x_2), (y_1, y_2) \in \mathcal{S}$. Then $L(x_1, x_2)L(y_1, y_2)$ contains only legitimate words iff $x_2 y_1 \in \{u_0, \epsilon\}$, in which case $L(x_1, x_2)L(y_1, y_2) \subseteq L(x_1, y_2)$. This motivates the following definition. For any $(x_1, x_2)$ and $(y_1, y_2)$ in $\mathcal{S}$, let

$$(x_1, x_2) \otimes (y_1, y_2) = \begin{cases} (x_1, y_2) & \text{if } x_2 y_1 \in \{u_0, \epsilon\} \\ \text{undefined} & \text{otherwise,} \end{cases}$$

so that $\otimes$ is a partial operation on $S$. Thus, if $(x_1, x_2) \otimes (y_1, y_2) = (z_1, z_2)$, then $L(x_1, x_2)L(y_1, y_2) \subseteq L(z_1, z_2)$, moreover, $(z_1, z_2)$ is the only element of $\mathcal{S}$ with this property.

Now let $(x_1, x_2) \in \mathcal{S}$ and consider a word $y$. Then $L(x_1, x_2)y$ contains only legitimate words iff $y \in L(y_1, y_2)$ for some $(y_1, y_2) \in \mathcal{S}$ such that $x_2 y_1 \in \{u_0, \epsilon\}$, in which case $(x_1, y_2)$ is the unique element of $\mathcal{S}$ with $L(x_1, x_2)y \subseteq L(x_1, y_2)$. Thus we define $(x_1, x_2) \otimes y = (x_1, y_2)$ if this holds, otherwise $(x_1, x_2) \otimes y$ is not defined. We define $y \otimes (x_1, x_2)$ symmetrically. The partial operation $\otimes$ is associative in a strong sense.

Using the above notions, the fourth condition of Theorem 3.1 can be rephrased as follows. For each strong component $\mathcal{C}$ containing a recursive nonterminal there is a primitive word $u_0 = u_0^{\mathcal{C}}$ (unique up to conjugacy) such that for all $X, Y \in \mathcal{C}$ there is (a necessarily unique) $(x_1, x_2) \in \mathcal{S}$ such that whenever $X \Rightarrow^+ wYp$ then $w \in L(x_1, x_2)$.

As before, let us assume that $G = (N, \{0, 1\}, P, S)$ is a context-free grammar that contains no useless nonterminals or $\epsilon$-rules. Moreover, we assume that $G$ is left-recursion free and that $L(G)$ is not empty.

**Lemma 4.1** *Suppose that the fourth condition of Theorem 3.1 holds and let $\mathcal{C}$ be a strong component containing a recursive nonterminal. Let $u_0 = u_0^{\mathcal{C}}$, and suppose that each nonterminal generates at least two terminal words. Then for each $X$ such that $X_0 \Rightarrow^* pXqYr$ for some $X_0, Y \in \mathcal{C}$ and words $p, q, r$ there is a unique $(x_1, x_2) \in \mathcal{S}$ with $L(X) \subseteq L(x_1, x_2)$.*

Proof. Let $(y_1, y_2)$ denote the unique element of $\mathcal{S}$ such that $w \in L(y_1, y_2)$ whenever $X_0 \Rightarrow^+ wYs$ for some $s$. Then $L(pXq) \subseteq L(y_1, y_2)$, so that $uL(X)v \subseteq L(y_1, y_2)$ for any fixed $u \in L(p)$ and $v \in L(q)$. This is possible only if $L(X) \subseteq L(x_1, x_2)$ for some $(x_1, x_2) \in \mathcal{S}$. Since $L(X)$ contains at least two words, $(x_1, x_2)$ is unique. □

**Theorem 4.2** *Suppose that each nonterminal generates a language of at least two words. Then $(L(G), <_\ell)$ is a scattered linear ordering iff the following holds for each strong component $\mathcal{C}$ containing a recursive nonterminal: There exists a primitive word $u_0$ such that for any two not necessarily different nonterminals $X$ and $Y$ in $\mathcal{C}$ there is some $\varphi(X,Y) \in \mathcal{S}$ and for each nonterminal $Z$ there is some $\psi(Z) \in \mathcal{S}$ such that*

$$\varphi(X,Y) \otimes \varphi(Y,Z) = \varphi(X,Z) \qquad (2)$$

*for all $X, Y, Z \in \mathcal{C}$, and such that the following hold for all productions $X \to w_0 Y_1 \ldots Y_k w_k$:*

1. *If $X \in \mathcal{C}$ and $Y_i \in \mathcal{C}$ for some $i$, then*

$$\varphi(X, Y_i) = w_0 \otimes \psi(Y_1) \otimes w_1 \otimes \ldots \otimes \psi(Y_{i-1}) \otimes w_{i-1}. \qquad (3)$$

2. *If there is derivation $X_0 \Rightarrow^* pXqYr$ for some $X_0, Y \in \mathcal{C}$, then*

$$\psi(X) = w_0 \otimes \psi(Y_1) \otimes \ldots \otimes \psi(Y_k) \otimes w_k. \qquad (4)$$

(In the degenerate case when $k = 0$ in the last equation, we mean that $w_0$ belongs to the language $L(\psi(X))$.

*Proof.* Suppose that the conditions of the Theorem hold. Consider a strong component $\mathcal{C}$ containing a recursive nonterminal and the corresponding primitive word $u_0$. Then for any $X$ such that there is derivation $X_0 \Rightarrow^* pXqYr$ for some $X_0, Y \in \mathcal{C}$ we have that $L(X) \subseteq L(\psi(X))$:

*Claim 1.* Suppose that (4) holds for all appropriate rules. Then for each $X$ such that there is a derivation $X_0 \Rightarrow^* pXqYr$ for some $X_0, Y \in \mathcal{C}$ it holds that $L(X) \subseteq L(\psi(X))$.

Indeed, suppose that $X \Rightarrow^* w$. We prove that $w \in L(\psi(X))$ by induction on the length of the derivation. When the length of the derivation is 1, the claim is clear by (4). Suppose that the length is greater than 1. Then there exist some rule $X \to w_0 Y_1 w_1 \ldots Y_k w_k$ and words $z_1, \ldots, z_k$ with $w = w_0 z_1 \ldots z_k w_k$ and $Y_i \Rightarrow^* z_i$ for all $i$. By the induction hypothesis we have that each $z_i$ is in $L(\psi(Y_i))$. Since (4) holds, we conclude that $w = w_0 z_1 \ldots z_k w_k \in L(\psi(X))$. This ends the proof of Claim 1.

Also, for any $X, Y \in \mathcal{C}$ and words $w$ and $p$ with $X \Rightarrow^+ wYp$ we have that $w \in L(\varphi(X,Y))$ as shown by the following claim:

*Claim 2.* Suppose that (2), (3) and (4) hold. Then if $X \Rightarrow^+ wYp$, where $X, Y \in \mathcal{C}$, then $w \in L(\varphi(X,Y))$.

To see this, consider a derivation tree whose root is labeled $X$ and whose frontier is $wYp$. Let $Y_1 = X, Y_2, \ldots, Y_\ell, Y_{\ell+1} = Y$ be all the nonterminal labels along the path from the root to the leaf labeled $Y$. Moreover, let $Y_i \to p_i Y_{i+1} q_i$ denote the rule used to rewrite $Y_i$, for $i = 1, \ldots, \ell$. By (3) we have that
$$\varphi(Y_1, Y_2) = \psi(p_1), \ldots, \varphi(Y_\ell, Y_{\ell+1}) = \psi(p_\ell)$$
where if $p_i = z_0 Z_1 \ldots Z_k z_k$, say, then $\psi(p_i) = z_0 \otimes \psi(Z_1) \otimes \ldots \otimes \psi(Z_k) \otimes z_{k+1}$. Now let us write $w = w_1 \ldots w_\ell$ with $p_i \Rightarrow^* w_i$ for all $i$. Using Claim 1 and the equality $\varphi(Y_i, Y_{i+1}) = \psi(p_i)$, we obtain $w_i \in L(\varphi(Y_i, Y_{i+1}))$. Since this holds for all $i$, we obtain by (2) that $w \in L(\varphi(X,Y))$.

We conclude that the fourth condition of Theorem 3.1 holds, so that $L(G)$ is scattered.

Suppose now that $L(G)$ is scattered. Then the fourth condition of Theorem 3.1 holds. Suppose that $\mathcal{C}$ is a strong component containing a recursive nonterminal. Let $u_0 = u_0^\mathcal{C}$. By assumption, for each $X, Y \in \mathcal{C}$ there exists a unique $(x_1, x_2) \in \mathcal{S}$ such that whenever $X \Rightarrow^+ wYp$ then $w \in L(x_1, x_2)$. Define $\varphi(X, Y) = (x_1, x_2)$. By Lemma 4.1, for each $X$ such that there is derivation $X_0 \Rightarrow^* pXqYr$ for some words $p, q, r$ and nonterminals $X_0, Y \in \mathcal{C}$, there is a unique $(x_1, x_2) \in \mathcal{S}$ with $L(X) \subseteq L(x_1, x_2)$. Define $\psi(X) = (x_1, x_2)$. The pairs so defined solve the system of equations in the Theorem. □

**Theorem 4.3** *It is decidable in exponential time whether a context-free language generated by a context-free grammar $G$ is scattered.*

*Proof.* Without loss of generality we may assume that the terminal alphabet is $\{0, 1\}$ and that the grammar $G$ contains no useless nonterminals or $\epsilon$-rules. Moreover, we may assume that $G$ is left-recursion free and each nonterminal generates at least two terminal words.

First, for each $\mathcal{C}$ containing a recursive nonterminal, one can compute in exponetial time a primitive word $u_0$ which is the only candidate for $u_0^\mathcal{C}$. This is done by finding in exponential time a left derivation $X \Rightarrow^+ wXp$, with $X \in \mathcal{C}$, then $u_0$ is the primitive root of $w$. Second, in the same way, for any $X, Y \in \mathcal{C}$, we can determine in exponential time the only candidate for $\varphi(X, Y)$ by computing a left derivation $X \Rightarrow^+ wYp$, where the length of $w$ is between $|u_0|$ and $2|u_0|$. Also, we can compute in exponential time the only candidate for $\psi(X)$, for all appropriate $X$. Then it remains to check that the equations of Theorem 4.2 hold. But there are a polynomial number of them, and the validity of each can be checked in exponential time. $\square$

The same result holds for deciding whether a context-free language is well-ordered.

**Theorem 4.4** *It is decidable in exponential time whether a context-free grammar generates a well-ordered language.*

*Proof.* Again, we may restrict the grammars as in the previous proof. The extra condition introduced in Theorem 3.7 can be tested in exponential time. Hint: if $X \Rightarrow_\ell^+ u_0^n u_1 Y p \Rightarrow u_0^n u1q$ is a left derivation, where $u0$ is a prefix of $u_0$, then the length of the derivation can be bounded by a exponential. $\square$

**Remark 4.5** *The algorithms given in the proofs of Theorem 4.3 and 4.4 run in polynomial time in the important special case when each nonterminal generates a prefix-free language of at least two words, since in that case whenever $X \to pYq$ is a rule such that $X \approx Y$, then $u \in \{0,1\}^*$.*

**Acknowledgement**

In a previous version of this paper, the algorithm presented in the last section was claimed to run in polynomial time. The author would like to thank Jean Berstel, Luc Boasson, Olivier Carton and Isabelle Fagnot for pointing out this error. As communicated by them, recently they have also found a proof of the fact that it is decidable for a context-free grammar whether it generates a scattered language.